\documentclass[conference,10pt,twocolumn]{IEEEtran}
\IEEEoverridecommandlockouts
\usepackage{cite}
\usepackage[T1]{fontenc}
\usepackage{graphicx}
\usepackage{amssymb}
\usepackage{amsmath}
\usepackage{subfigure}
\usepackage{amsthm}
\usepackage{microtype} 
\usepackage{balance}
\usepackage{color}

\usepackage{amssymb}
\usepackage{amsfonts}
\usepackage{mathrsfs}
\usepackage{xspace}
\usepackage{bm}
\usepackage{upgreek}

\newcommand{\safemath}[2]{\newcommand{#1}{\ensuremath{#2}\xspace}}

\safemath{\bma}{\mathbf{a}}
\safemath{\bmb}{\mathbf{b}}
\safemath{\bmc}{\mathbf{c}}
\safemath{\bmd}{\mathbf{d}}
\safemath{\bme}{\mathbf{e}}
\safemath{\bmf}{\mathbf{f}}
\safemath{\bmg}{\mathbf{g}}
\safemath{\bmh}{\mathbf{h}}
\safemath{\bmi}{\mathbf{i}}
\safemath{\bmj}{\mathbf{j}}
\safemath{\bmk}{\mathbf{k}}
\safemath{\bml}{\mathbf{l}}
\safemath{\bmm}{\mathbf{m}}
\safemath{\bmn}{\mathbf{n}}
\safemath{\bmo}{\mathbf{o}}
\safemath{\bmp}{\mathbf{p}}
\safemath{\bmq}{\mathbf{q}}
\safemath{\bmr}{\mathbf{r}}
\safemath{\bms}{\mathbf{s}}
\safemath{\bmt}{\mathbf{t}}
\safemath{\bmu}{\mathbf{u}}
\safemath{\bmv}{\mathbf{v}}
\safemath{\bmw}{\mathbf{w}}
\safemath{\bmx}{\mathbf{x}}
\safemath{\bmy}{\mathbf{y}}
\safemath{\bmz}{\mathbf{z}}
\safemath{\bmzero}{\mathbf{0}}
\safemath{\bmone}{\mathbf{1}}

\bmdefine{\biad}{a}
\bmdefine{\bibd}{b}
\bmdefine{\bicd}{c}
\bmdefine{\bidd}{d}
\bmdefine{\bied}{e}
\bmdefine{\bifd}{f}
\bmdefine{\bigd}{g}
\bmdefine{\bihd}{h}
\bmdefine{\biid}{i}
\bmdefine{\bijd}{j}
\bmdefine{\bikd}{k}
\bmdefine{\bild}{l}
\bmdefine{\bimd}{m}
\bmdefine{\bind}{n}
\bmdefine{\biod}{o}
\bmdefine{\bipd}{p}
\bmdefine{\biqd}{q}
\bmdefine{\bird}{r}
\bmdefine{\bisd}{s}
\bmdefine{\bitd}{t}
\bmdefine{\biud}{u}
\bmdefine{\bivd}{v}
\bmdefine{\biwd}{w}
\bmdefine{\bixd}{x}
\bmdefine{\biyd}{y}
\bmdefine{\bizd}{z}

\bmdefine{\bixid}{\xi}
\bmdefine{\bilambdad}{\lambda}
\bmdefine{\bimud}{\mu}
\bmdefine{\bithetad}{\theta}
\bmdefine{\biphid}{\phi}
\bmdefine{\bideltad}{\delta}

\safemath{\bmia}{\biad}
\safemath{\bmib}{\bibd}
\safemath{\bmic}{\bicd}
\safemath{\bmid}{\bidd}
\safemath{\bmie}{\bied}
\safemath{\bmif}{\bifd}
\safemath{\bmig}{\bigd}
\safemath{\bmih}{\bihd}
\safemath{\bmii}{\biid}
\safemath{\bmij}{\bijd}
\safemath{\bmik}{\bikd}
\safemath{\bmil}{\bild}
\safemath{\bmim}{\bimd}
\safemath{\bmin}{\bind}
\safemath{\bmio}{\biod}
\safemath{\bmip}{\bipd}
\safemath{\bmiq}{\biqd}
\safemath{\bmir}{\bird}
\safemath{\bmis}{\bisd}
\safemath{\bmit}{\bitd}
\safemath{\bmiu}{\biud}
\safemath{\bmiv}{\bivd}
\safemath{\bmiw}{\biwd}
\safemath{\bmix}{\bixd}
\safemath{\bmiy}{\biyd}
\safemath{\bmiz}{\bizd}

\safemath{\bmxi}{\bixid}
\safemath{\bmlambda}{\bilambdad}
\safemath{\bmmu}{\bimud}
\safemath{\bmtheta}{\bithetad}
\safemath{\bmphi}{\biphid}
\safemath{\bmdelta}{\bideltad}

\safemath{\bA}{\mathbf{A}}
\safemath{\bB}{\mathbf{B}}
\safemath{\bC}{\mathbf{C}}
\safemath{\bD}{\mathbf{D}}
\safemath{\bE}{\mathbf{E}}
\safemath{\bF}{\mathbf{F}}
\safemath{\bG}{\mathbf{G}}
\safemath{\bH}{\mathbf{H}}
\safemath{\bI}{\mathbf{I}}
\safemath{\bJ}{\mathbf{J}}
\safemath{\bK}{\mathbf{K}}
\safemath{\bL}{\mathbf{L}}
\safemath{\bM}{\mathbf{M}}
\safemath{\bN}{\mathbf{N}}
\safemath{\bO}{\mathbf{O}}
\safemath{\bP}{\mathbf{P}}
\safemath{\bQ}{\mathbf{Q}}
\safemath{\bR}{\mathbf{R}}
\safemath{\bS}{\mathbf{S}}
\safemath{\bT}{\mathbf{T}}
\safemath{\bU}{\mathbf{U}}
\safemath{\bV}{\mathbf{V}}
\safemath{\bW}{\mathbf{W}}
\safemath{\bX}{\mathbf{X}}
\safemath{\bY}{\mathbf{Y}}
\safemath{\bZ}{\mathbf{Z}}

\safemath{\bDelta}{\mathbf{\Delta}}
\safemath{\bLambda}{\mathbf{\UpLambda}}
\safemath{\bPhi}{\mathbf{\Upphi}}
\safemath{\bSigma}{\mathbf{\Upsigma}}
\safemath{\bOmega}{\mathbf{\Upomega}}
\safemath{\bTheta}{\mathbf{\Uptheta}}

\bmdefine{\biAd}{A}
\bmdefine{\biBd}{B}
\bmdefine{\biCd}{C}
\bmdefine{\biDd}{D}
\bmdefine{\biEd}{E}
\bmdefine{\biFd}{F}
\bmdefine{\biGd}{G}
\bmdefine{\biHd}{H}
\bmdefine{\biId}{I}
\bmdefine{\biJd}{J}
\bmdefine{\biKd}{K}
\bmdefine{\biLd}{L}
\bmdefine{\biMd}{M}
\bmdefine{\biOd}{N}
\bmdefine{\biPd}{O}
\bmdefine{\biQd}{P}
\bmdefine{\biRd}{R}
\bmdefine{\biSd}{S}
\bmdefine{\biTd}{T}
\bmdefine{\biUd}{U}
\bmdefine{\biVd}{V}
\bmdefine{\biWd}{W}
\bmdefine{\biXd}{X}
\bmdefine{\biYd}{Y}
\bmdefine{\biZd}{Z}

\bmdefine{\biDelta}{\Delta}
\bmdefine{\biLambda}{\Lambda}
\bmdefine{\biPhi}{\Phi}
\bmdefine{\biSigma}{\Sigma}
\bmdefine{\biOmega}{\Omega}
\bmdefine{\biTheta}{\Theta}

\safemath{\bimA}{\biAd}
\safemath{\bimB}{\biBd}
\safemath{\bimC}{\biCd}
\safemath{\bimD}{\biDd}
\safemath{\bimE}{\biEd}
\safemath{\bimF}{\biFd}
\safemath{\bimG}{\biGd}
\safemath{\bimH}{\biHd}
\safemath{\bimI}{\biId}
\safemath{\bimJ}{\biJd}
\safemath{\bimK}{\biKd}
\safemath{\bimL}{\biLd}
\safemath{\bimM}{\biMd}
\safemath{\bimN}{\biNd}
\safemath{\bimO}{\biOd}
\safemath{\bimP}{\biPd}
\safemath{\bimQ}{\biQd}
\safemath{\bimR}{\biRd}
\safemath{\bimS}{\biSd}
\safemath{\bimT}{\biTd}
\safemath{\bimU}{\biUd}
\safemath{\bimV}{\biVd}
\safemath{\bimW}{\biWd}
\safemath{\bimX}{\biXd}
\safemath{\bimY}{\biYd}
\safemath{\bimZ}{\biZd}

\safemath{\bimDelta}{\biDelta}
\safemath{\bimLambda}{\biLambda}
\safemath{\bimPhi}{\biPhi}
\safemath{\bimSigma}{\biSigma}
\safemath{\bimOmega}{\biOmega}
\safemath{\bimTheta}{\biTheta}

\safemath{\setA}{\mathcal{A}}
\safemath{\setB}{\mathcal{B}}
\safemath{\setC}{\mathcal{C}}
\safemath{\setD}{\mathcal{D}}
\safemath{\setE}{\mathcal{E}}
\safemath{\setF}{\mathcal{F}}
\safemath{\setG}{\mathcal{G}}
\safemath{\setH}{\mathcal{H}}
\safemath{\setI}{\mathcal{I}}
\safemath{\setJ}{\mathcal{J}}
\safemath{\setK}{\mathcal{K}}
\safemath{\setL}{\mathcal{L}}
\safemath{\setM}{\mathcal{M}}
\safemath{\setN}{\mathcal{N}}
\safemath{\setO}{\mathcal{O}}
\safemath{\setP}{\mathcal{P}}
\safemath{\setQ}{\mathcal{Q}}
\safemath{\setR}{\mathcal{R}}
\safemath{\setS}{\mathcal{S}}
\safemath{\setT}{\mathcal{T}}
\safemath{\setU}{\mathcal{U}}
\safemath{\setV}{\mathcal{V}}
\safemath{\setW}{\mathcal{W}}
\safemath{\setX}{\mathcal{X}}
\safemath{\setY}{\mathcal{Y}}
\safemath{\setZ}{\mathcal{Z}}
\safemath{\emptySet}{\varnothing}

\safemath{\colA}{\mathscr{A}}
\safemath{\colB}{\mathscr{B}}
\safemath{\colC}{\mathscr{C}}
\safemath{\colD}{\mathscr{D}}
\safemath{\colE}{\mathscr{E}}
\safemath{\colF}{\mathscr{F}}
\safemath{\colG}{\mathscr{G}}
\safemath{\colH}{\mathscr{H}}
\safemath{\colI}{\mathscr{I}}
\safemath{\colJ}{\mathscr{J}}
\safemath{\colK}{\mathscr{K}}
\safemath{\colL}{\mathscr{L}}
\safemath{\colM}{\mathscr{M}}
\safemath{\colN}{\mathscr{N}}
\safemath{\colO}{\mathscr{O}}
\safemath{\colP}{\mathscr{P}}
\safemath{\colQ}{\mathscr{Q}}
\safemath{\colR}{\mathscr{R}}
\safemath{\colS}{\mathscr{S}}
\safemath{\colT}{\mathscr{T}}
\safemath{\colU}{\mathscr{U}}
\safemath{\colV}{\mathscr{V}}
\safemath{\colW}{\mathscr{W}}
\safemath{\colX}{\mathscr{X}}
\safemath{\colY}{\mathscr{Y}}
\safemath{\colZ}{\mathscr{Z}}

\safemath{\opA}{\mathbb{A}}
\safemath{\opB}{\mathbb{B}}
\safemath{\opC}{\mathbb{C}}
\safemath{\opD}{\mathbb{D}}
\safemath{\opE}{\mathbb{E}}
\safemath{\opF}{\mathbb{F}}
\safemath{\opG}{\mathbb{G}}
\safemath{\opH}{\mathbb{H}}
\safemath{\opI}{\mathbb{I}}
\safemath{\opJ}{\mathbb{J}}
\safemath{\opK}{\mathbb{K}}
\safemath{\opL}{\mathbb{L}}
\safemath{\opM}{\mathbb{M}}
\safemath{\opN}{\mathbb{N}}
\safemath{\opO}{\mathbb{O}}
\safemath{\opP}{\mathbb{P}}
\safemath{\opQ}{\mathbb{Q}}
\safemath{\opR}{\mathbb{R}}
\safemath{\opS}{\mathbb{S}}
\safemath{\opT}{\mathbb{T}}
\safemath{\opU}{\mathbb{U}}
\safemath{\opV}{\mathbb{V}}
\safemath{\opW}{\mathbb{W}}
\safemath{\opX}{\mathbb{X}}
\safemath{\opY}{\mathbb{Y}}
\safemath{\opZ}{\mathbb{Z}}
\safemath{\opZero}{\mathbb{O}}
\safemath{\identityop}{\opI}

\safemath{\veca}{\bma}
\safemath{\vecb}{\bmb}
\safemath{\vecc}{\bmc}
\safemath{\vecd}{\bmd}
\safemath{\vece}{\bme}
\safemath{\vecf}{\bmf}
\safemath{\vecg}{\bmg}
\safemath{\vech}{\bmh}
\safemath{\veci}{\bmi}
\safemath{\vecj}{\bmj}
\safemath{\veck}{\bmk}
\safemath{\vecl}{\bml}
\safemath{\vecm}{\bmm}
\safemath{\vecn}{\bmn}
\safemath{\veco}{\bmo}
\safemath{\vecp}{\bmp}
\safemath{\vecq}{\bmq}
\safemath{\vecr}{\bmr}
\safemath{\vecs}{\bms}
\safemath{\vect}{\bmt}
\safemath{\vecu}{\bmu}
\safemath{\vecv}{\bmv}
\safemath{\vecw}{\bmw}
\safemath{\vecx}{\bmx}
\safemath{\vecy}{\bmy}
\safemath{\vecz}{\bmz}

\safemath{\veczero}{\bmzero}
\safemath{\vecone}{\bmone}
\safemath{\vecxi}{\bmxi}
\safemath{\veclambda}{\bmlambda}
\safemath{\vecmu}{\bmmu}
\safemath{\vectheta}{\bmtheta}
\safemath{\vecphi}{\bmphi}
\safemath{\vecdelta}{\bmdelta}

\safemath{\matA}{\bA}
\safemath{\matB}{\bB}
\safemath{\matC}{\bC}
\safemath{\matD}{\bD}
\safemath{\matE}{\bE}
\safemath{\matF}{\bF}
\safemath{\matG}{\bG}
\safemath{\matH}{\bH}
\safemath{\matI}{\bI}
\safemath{\matJ}{\bJ}
\safemath{\matK}{\bK}
\safemath{\matL}{\bL}
\safemath{\matM}{\bM}
\safemath{\matN}{\bN}
\safemath{\matO}{\bO}
\safemath{\matP}{\bP}
\safemath{\matQ}{\bQ}
\safemath{\matR}{\bR}
\safemath{\matS}{\bS}
\safemath{\matT}{\bT}
\safemath{\matU}{\bU}
\safemath{\matV}{\bV}
\safemath{\matW}{\bW}
\safemath{\matX}{\bX}
\safemath{\matY}{\bY}
\safemath{\matZ}{\bZ}
\safemath{\matzero}{\bmzero}

\safemath{\matDelta}{\bDelta}
\safemath{\matLambda}{\bLambda}
\safemath{\matPhi}{\bPhi}
\safemath{\matSigma}{\bSigma}
\safemath{\matOmega}{\bOmega}
\safemath{\matTheta}{\bTheta}

\safemath{\matidentity}{\matI}
\safemath{\matone}{\matO}

\safemath{\rnda}{A}
\safemath{\rndb}{B}
\safemath{\rndc}{C}
\safemath{\rndd}{D}
\safemath{\rnde}{E}
\safemath{\rndf}{F}
\safemath{\rndg}{G}
\safemath{\rndh}{H}
\safemath{\rndi}{I}
\safemath{\rndj}{J}
\safemath{\rndk}{K}
\safemath{\rndl}{L}
\safemath{\rndm}{M}
\safemath{\rndn}{N}
\safemath{\rndo}{O}
\safemath{\rndp}{P}
\safemath{\rndq}{Q}
\safemath{\rndr}{R}
\safemath{\rnds}{S}
\safemath{\rndt}{T}
\safemath{\rndu}{U}
\safemath{\rndv}{V}
\safemath{\rndw}{W}
\safemath{\rndx}{X}
\safemath{\rndy}{Y}
\safemath{\rndz}{Z}

\safemath{\rveca}{\bimA}
\safemath{\rvecb}{\bimB}
\safemath{\rvecc}{\bimC}
\safemath{\rvecd}{\bimD}
\safemath{\rvece}{\bimE}
\safemath{\rvecf}{\bimF}
\safemath{\rvecg}{\bimG}
\safemath{\rvech}{\bimH}
\safemath{\rveci}{\bimI}
\safemath{\rvecj}{\bimJ}
\safemath{\rveck}{\bimK}
\safemath{\rvecl}{\bimL}
\safemath{\rvecm}{\bimM}
\safemath{\rvecn}{\bimN}
\safemath{\rveco}{\bomO}
\safemath{\rvecp}{\bimP}
\safemath{\rvecq}{\bimQ}
\safemath{\rvecr}{\bimR}
\safemath{\rvecs}{\bimS}
\safemath{\rvect}{\bimT}
\safemath{\rvecu}{\bimU}
\safemath{\rvecv}{\bimV}
\safemath{\rvecw}{\bimW}
\safemath{\rvecx}{\bimX}
\safemath{\rvecy}{\bimY}
\safemath{\rvecz}{\bimZ}

\safemath{\rvecxi}{\bmxi}
\safemath{\rveclambda}{\bmlambda}
\safemath{\rvecmu}{\bmmu}
\safemath{\rvectheta}{\bmtheta}
\safemath{\rvecphi}{\bmphi}

\safemath{\rmatA}{\bimA}
\safemath{\rmatB}{\bimB}
\safemath{\rmatC}{\bimC}
\safemath{\rmatD}{\bimD}
\safemath{\rmatE}{\bimE}
\safemath{\rmatF}{\bimF}
\safemath{\rmatG}{\bimG}
\safemath{\rmatH}{\bimH}
\safemath{\rmatI}{\bimI}
\safemath{\rmatJ}{\bimJ}
\safemath{\rmatK}{\bimK}
\safemath{\rmatL}{\bimL}
\safemath{\rmatM}{\bimM}
\safemath{\rmatN}{\bimN}
\safemath{\rmatO}{\bimO}
\safemath{\rmatP}{\bimP}
\safemath{\rmatQ}{\bimQ}
\safemath{\rmatR}{\bimR}
\safemath{\rmatS}{\bimS}
\safemath{\rmatT}{\bimT}
\safemath{\rmatU}{\bimU}
\safemath{\rmatV}{\bimV}
\safemath{\rmatW}{\bimW}
\safemath{\rmatX}{\bimX}
\safemath{\rmatY}{\bimY}
\safemath{\rmatZ}{\bimZ}

\safemath{\rmatDelta}{\bimDelta}
\safemath{\rmatLambda}{\bimLambda}
\safemath{\rmatPhi}{\bimPhi}
\safemath{\rmatSigma}{\bimSigma}
\safemath{\rmatOmega}{\bimOmega}
\safemath{\rmatTheta}{\bimTheta}

\usepackage{amssymb}
\usepackage{amsfonts}
\usepackage{mathrsfs}
\usepackage{xspace}
\usepackage{bm}
\usepackage{fancyref}
\usepackage{textcomp}

\usepackage{multirow}
\usepackage{stmaryrd}

\newenvironment{textbmatrix}{	\setlength{\arraycolsep}{2.5pt}%
								\big[\begin{matrix}}{\end{matrix}\big]%
								\raisebox{0.08ex}{\vphantom{M}}}

\def\be{\begin{equation}}
\def\ee{\end{equation}}
\def\een{\nonumber \end{equation}}
\def\mat{\begin{bmatrix}}
\def\emat{\end{bmatrix}}
\def\btm{\begin{textbmatrix}}
\def\etm{\end{textbmatrix}}

\def\ba#1\ea{\begin{align}#1\end{align}}
\def\bas#1\eas{\begin{align*}#1\end{align*}}
\def\bs#1\es{\begin{split}#1\end{split}}
\def\bg#1\eg{\begin{gather}#1\end{gather}}
\def\bml#1\eml{\begin{multline}#1\end{multline}}
\def\bi#1\ei{\begin{itemize}#1\end{itemize}}

\newcommand{\lefto}{\mathopen{}\left}

\DeclareMathOperator{\Tr}{\opT r}			
\DeclareMathOperator*{\argmin}{arg\;min}		
\DeclareMathOperator*{\argmax}{arg\;max}		
\DeclareMathOperator{\Exop}{\opE}			

\newcommand{\abs}[1]{\lefto\lvert#1\right\rvert}		

\newcommand{\vecnorm}[1]{\lefto\lVert#1\right\rVert}		

\safemath{\dirac}{\delta}					
\safemath{\krond}{\dirac}					

\safemath{\upto}{\uparrow}
\safemath{\downto}{\downarrow}
\safemath{\iu}{j}							
\safemath{\ev}{\lambda}						
\safemath{\hilseqspace}{l^{2}}				
\newcommand{\banachfunspace}[1]{\setL^{#1}}	
\safemath{\hilfunspace}{\banachfunspace{2}}	

\safemath{\SNR}{\textit{SNR}} 				
\safemath{\PAR}{\textit{PAR}} 				
\safemath{\No}{N_0}							
\safemath{\Es}{E_s}							
\safemath{\Eb}{E_b}							
\safemath{\EbNo}{\frac{\Eb}{\No}}
\safemath{\EsNo}{\frac{\Es}{\No}}

\DeclareMathOperator{\CHop}{\ensuremath{\opH}} 
\safemath{\tvir}{\rndh_{\CHop}}				
\safemath{\tvtf}{\rndl_{\CHop}}				
\safemath{\spf}{\rnds_{\CHop}}				
\safemath{\bff}{H_{\CHop}}					

\safemath{\ircf}{r_{h}}						
\safemath{\tftvcf}{r_{s}}					
\safemath{\tfcf}{r_{l}}						
\safemath{\bfcf}{r_{H}}						

\safemath{\tcorr}{c_h}						
\safemath{\scf}{c_{s}}						
\safemath{\tfcorr}{c_{l}}					
\safemath{\fcorr}{c_{H}}						

\safemath{\mi}{I}							
\safemath{\capacity}{C}						

\safemath{\normal}{\mathcal{N}}			
\safemath{\jpg}{\mathcal{CN}}			
\safemath{\mchain}{\leftrightarrow}		

\safemath{\dB}{\,\mathrm{dB}}
\safemath{\dBm}{\,\mathrm{dBm}}
\safemath{\Hz}{\,\mathrm{Hz}}
\safemath{\kHz}{\,\mathrm{kHz}}
\safemath{\MHz}{\,\mathrm{MHz}}
\safemath{\GHz}{\,\mathrm{GHz}}
\safemath{\s}{\,\mathrm{s}}
\safemath{\ms}{\,\mathrm{ms}}
\safemath{\mus}{\,\mathrm{\text{\textmu}s}}
\safemath{\ns}{\,\mathrm{ns}}
\safemath{\ps}{\,\mathrm{ps}}
\safemath{\meter}{\,\mathrm{m}}
\safemath{\mm}{\,\mathrm{mm}}
\safemath{\cm}{\,\mathrm{cm}}
\safemath{\m}{\,\mathrm{m}}
\safemath{\W}{\,\mathrm{W}}
\safemath{\mW}{\, \mathrm{mW}}
\safemath{\J}{\,\mathrm{J}}
\safemath{\K}{\,\mathrm{K}}
\safemath{\bit}{\,\mathrm{bit}}
\safemath{\nat}{\,\mathrm{nat}}

\safemath{\define}{\triangleq}			

\safemath{\equivalent}{\sim}
\safemath{\distas}{\sim}					
\safemath{\sdiff}{\Delta}				

\safemath{\reals}{\mathbb{R}}
\safemath{\positivereals}{\reals_{+}}
\safemath{\integers}{\mathbb{Z}}
\safemath{\posint}{\integers_{+}}
\safemath{\naturals}{\mathbb{N}}
\safemath{\posnaturals}{\naturals_{+}}
\safemath{\complexset}{\mathbb{C}}
\safemath{\rationals}{\mathbb{Q}}

\newcommand*{\fancyrefapplabelprefix}{app}		
\newcommand*{\fancyrefthmlabelprefix}{thm}		
\newcommand*{\fancyreflemlabelprefix}{lem}		
\newcommand*{\fancyrefcorlabelprefix}{cor}		
\newcommand*{\fancyrefdeflabelprefix}{def}		
\newcommand*{\fancyrefproplabelprefix}{prop}		
\newcommand*{\fancyrefexmpllabelprefix}{exmpl}
\newcommand*{\fancyrefalglabelprefix}{alg}		
\newcommand*{\fancyreftbllabelprefix}{tbl}		

\frefformat{vario}{\fancyrefseclabelprefix}{Section~#1}
\frefformat{vario}{\fancyrefthmlabelprefix}{Theorem~#1}
\frefformat{vario}{\fancyreftbllabelprefix}{Table~#1}
\frefformat{vario}{\fancyreflemlabelprefix}{Lemma~#1}
\frefformat{vario}{\fancyrefcorlabelprefix}{Corollary~#1}
\frefformat{vario}{\fancyrefdeflabelprefix}{Definition~#1}
\frefformat{vario}{\fancyreffiglabelprefix}{Fig.~#1}
\frefformat{vario}{\fancyrefapplabelprefix}{Appendix~#1}
\frefformat{vario}{\fancyrefeqlabelprefix}{(#1)}
\frefformat{vario}{\fancyrefproplabelprefix}{Proposition~#1}
\frefformat{vario}{\fancyrefexmpllabelprefix}{Example~#1}
\frefformat{vario}{\fancyrefalglabelprefix}{Algorithm~#1}

\safemath{\dictab}{[\,\dicta\,\,\dictb\,]}

\safemath{\ysig}{\bmy}
\safemath{\ysighat}{\hat{\ysig}}
\safemath{\ysigdim}{M}
\safemath{\xsig}{\bmx}
\safemath{\xsigdim}{N}
\safemath{\nx}{n_x}
\safemath{\zsig}{\bmz}
\safemath{\zsigdim}{\ysigdim}
\safemath{\rsig}{\bmr}
\safemath{\Adict}{\bA}
\safemath{\Adicttilde}{\widetilde{\Adict}}
\safemath{\Adictdim}{\outputdim\times\xsigdim}
\safemath{\avec}{\bma}
\safemath{\avectilde}{\tilde{\avec}}
\safemath{\Bdict}{\bB}
\safemath{\Bdicttilde}{\widetilde{\Bdict}}
\safemath{\Cdict}{\bC}
\safemath{\cvec}{\bmc}
\safemath{\Ddict}{\bD}
\safemath{\Ddictdim}{\ysigdim\times\xsigdim}
\safemath{\dvec}{\bmd}
\safemath{\Ddicttilde}{\widetilde{\bD}}
\safemath{\Bonb}{\bB}
\safemath{\bvec}{\bmb}
\safemath{\Bonbdim}{\ysigdim\times\ysigdim}
\safemath{\noise}{\bmn}
\safemath{\noisedim}{\ysigim}
\safemath{\err}{\bme}
\safemath{\errdim}{\ysigdim}
\safemath{\errset}{\setE}
\safemath{\nerr}{n_e}
\safemath{\delop}{\bP_\errset}
\safemath{\delopc}{\bP_{{\errset}^c}}

\safemath{\cplxi}{\imath}
\safemath{\cplxj}{\jmath}

\safemath{\dict}{\matD}
\safemath{\inputdim}{N}		
\safemath{\outputdim}{M}		
\safemath{\sparsity}{S}	
\safemath{\inputdimA}{{N_a}}	
\safemath{\inputdimB}{{N_b}}	
\safemath{\elemA}{{n_a}}	
\safemath{\elemB}{{n_b}}	
\safemath{\resA}{\matR_a}	
\safemath{\resB}{\matR_b}	
\safemath{\subD}{\matS} 
\safemath{\subA}{\matS_a} 
\safemath{\subB}{\matS_b} 
\safemath{\dicta}{\matA} 	
\safemath{\dictb}{\matB} 	
\safemath{\hollowS}{H}
\safemath{\hollowA}{H_a}
\safemath{\hollowB}{H_b}
\safemath{\cross}{Z}
\safemath{\coh}{\mu_d}			
\safemath{\coha}{\mu_a}			
\safemath{\cohb}{\mu_b}			
\safemath{\mubs}{\nu}	
\safemath{\cohm}{\mu_m} 
\safemath{\dictset}{\setD}	
\safemath{\dictsetp}{\dictset(\coh,\coha,\cohb)}	
\safemath{\dictsetgen}{\dictset_\text{gen}}
\safemath{\dictsetgenp}{\dictsetgen(\coh)}
\safemath{\dictsetonb}{\dictset_\text{onb}}
\safemath{\dictsetonbp}{\dictsetonb(\coh)}

\safemath{\leftside}{U}
\safemath{\rightsideA}{R_a}
\safemath{\rightsideB}{R_b}

\safemath{\indexS}{\setI_S} 

\safemath{\na}{n_a}			
\safemath{\nb}{n_b}			
\safemath{\coeffa}{p_i}	
\safemath{\coeffb}{q_j}	
\safemath{\seta}{\setP}		
\safemath{\setb}{\setQ}     
\safemath{\setw}{\setW}	
\safemath{\setz}{\setZ}	
\safemath{\cola}{\veca}		
\safemath{\colb}{\vecb}		
\safemath{\cold}{\vecd}		
\safemath{\inputvec}{\vecx} 	
\safemath{\error}{\vece}	
\safemath{\noiseout}{\vecz} 	
\safemath{\inputvecel}{x}
\safemath{\inputveca}{\vecx_a}
\safemath{\inputvecb}{\vecx_b}
\safemath{\outputvec}{\vecy}	
\safemath{\lambdamin}{\lambda_{\mathrm{min}}}

\newcommand{\normtwo}[1]{\vecnorm{#1}_2}

\newcommand{\normfro}[1]{\vecnorm{#1}_F}

\safemath{\elltwo}{\ell_2}
\safemath{\ellone}{\ell_1}
\safemath{\ellzero}{\ell_0}
\safemath{\ellinf}{\ell_\infty}
\safemath{\ellinftilde}{\ell_{\widetilde\infty}}
\safemath{\licard}{Z(\coh,\coha,\cohb)}
\safemath{\xsol}{\hat{x}}
\safemath{\xbord}{x_b}		
\safemath{\xstat}{x_s}		
\safemath{\xstatLone}{\tilde{x}_s}
\safemath{\order}{\mathcal{O}} 
\safemath{\scales}{\Theta} 
\safemath{\ones}{\mathbf{1}} 
\safemath{\zeroes}{\mathbf{0}} 
\safemath{\thlone}{\kappa(\coh,\cohb)} 
\safemath{\constoneA}{\delta} 
\safemath{\constoneB}{\epsilon} 
\safemath{\nlarge}{L}				   
\safemath{\sumlarge}{S_\nlarge}
\safemath{\maxlarger}{P_\nlarge}	   
\safemath{\Pzero}{\textrm{P0}}	
\safemath{\Pone}{\textrm{P1}}
\safemath{\vecfir}{\vecw}			 
\safemath{\vecsec}{\vecz}
\safemath{\elvecfir}{w}              
\safemath{\elvecsec}{z}				 
\safemath{\nlargefir}{n}
\safemath{\normout}{\gamma}
\safemath{\auxfun}{h}
\safemath{\supp}{\textrm{supp}}

\safemath{\indexa}{\ell}
\safemath{\indexb}{r}
\safemath{\indexc}{i}
\safemath{\indexd}{j}

\safemath{\project}{P}

\allowdisplaybreaks

\definecolor{Midnightblue}{rgb}{0.1, 0.1, 0.44}

\begin{document}
	
\title{Soft-Output Joint Channel Estimation and \\ Data Detection using Deep Unfolding}

\author{
	\IEEEauthorblockN{Haochuan Song$^{1,2}$, Xiaohu You$^{1,2}$, Chuan Zhang$^{1,2}$, and Christoph Studer$^3$} \\

	\IEEEauthorblockA{ \em 
		$^1$National Mobile Communications Research Laboratory, Southeast University, Nanjing, China\\
		$^2$Purple Mountain Laboratories, Nanjing, China\\
		$^3$Department of Information Technology and Electrical Engineering, ETH Z\"urich, Z\"urich, Switzerland\\
		e-mail: \{hcsong, xhyu, chzhang\}@seu.edu.cn;  studer@ethz.ch
		}
		\\[-0.99cm]
		
		\thanks{
		The work of HS, XY, and CZ was supported in part by the National Key R\&D Program of China under Grant 2020YFB2205503, by the NSFC under Grants 61871115 and 62122020, and by the Jiangsu NSF under Grant BK20211512. The work of CS was supported in part by ComSenTer, one of six centers in JUMP, a SRC program sponsored by DARPA, by an ETH Research Grant, and by the US NSF under grants CNS-1717559 and ECCS-1824379.}
\thanks{The authors thank Gian Marti for comments and suggestions. 
		}
}

\maketitle
\begin{abstract}
We propose a novel soft-output joint channel estimation and data detection (JED) algorithm for multiuser (MU) multiple-input multiple-output (MIMO) wireless communication systems.
Our algorithm approximately solves a maximum a-posteriori JED optimization problem using deep unfolding and generates soft-output information for the transmitted bits in every iteration. 
The parameters of the unfolded algorithm are computed by a hyper-network that is trained with a binary cross entropy (BCE) loss. 
We evaluate the performance of our algorithm in a coded MU-MIMO system with $\bf 8$ basestation antennas and $\bf 4$ user equipments and compare it to state-of-the-art algorithms separate channel estimation from soft-output data detection. 
Our results demonstrate that our JED algorithm outperforms such data detectors with as few as 10 iterations. 
\end{abstract}

\section{Introduction}
In the uplink of multiuser multiple-input multiple-output (MU-MIMO) systems, where user equipments (UEs) transmit pilots and data to a base station (BS), deploying optimal joint channel estimation and data detection (JED) is an elusive goal, and has been practicable only for small-scale MIMO systems due to the combinatorial nature and the high-dimensionality of the JED problem~\cite{vikalo2006efficient,alshamary2015optimal,xu2008exact}.
To overcome the high complexity of such methods, references \cite{castanedaData2016,castaneda2017vlsi} proposed gradient-based algorithms that efficiently compute approximate solutions to the single-UE JED problem for large BS antenna arrays. 
The multi-UE JED case has been tackled recently in~\cite{songMinimizing2020} for cell-free massive MU-MIMO systems, also using a gradient-based algorithm. 
All of these methods compute hard-output estimates and are, thus, unable to realize the full potential of coded data transmission.
A soft-output algorithm that approximates JED using iterative estimation and detection (IED) has been proposed recently in \cite{heModeldriven2020}, which alternates between channel estimation and data detection. Similar alternating optimization methods have been used before for IED  in~\cite{Yilmaz19a,kofidis2017joint}. 

\subsection{Contributions}

With the recent progress in deep neural networks, optimal JED~\cite{vikalo2006efficient,alshamary2015optimal,xu2008exact} is suddenly within reach by leveraging deep unfolding of iterative algorithms \cite{hersheyDeep2014,balatsoukas-stimming19a,mongaAlgorithm2021,goutayDeep2020,heModeldriven2020}. 
In this paper, we propose a novel soft-output JED algorithm that  builds upon a maximum a-posteriori (MAP) JED problem formulation which we solve approximately using deep unfolding. 
We derive an iterative algorithm with soft-output capabilities by utilizing the approximate posterior mean estimator (PME) put forward in~\cite{jeonMismatched2020,jeon3542019}.
All algorithm parameters are generated by a hyper-network that processes estimated channel state information (CSI). The hyper-network is trained with a binary cross entropy (BCE) loss that exploits the soft-output capabilities of our JED algorithm.
We provide simulation results for an $8$ BS antenna, $4$ UE MU-MIMO system and compare our algorithm to state-of-the-art methods that separate channel estimation from soft-output data detection. 

\subsection{Notation}
Lower case letters denote matrices and upper case boldface letters denote vectors. We use $ A_{b,u} $,  $\bma_u$, and $ b_k $ to represent the entry in the $ b $th row and $ u $th column of the matrix $ \bA $, the $u$th column of matrix $\bA$, and the $ k $th entry in the vector~$ \bmb $, respectively. 
The superscripts $^T $ and $^H$ denote the transpose and Hermitian transpose, respectively. The Frobenius norm and trace of a matrix $\bA$ is $ \normfro{\bA} $ and $\Tr(\bA)$. The $U\times U$ identity matrix is $\bI_U$. 
Sets are denoted by calligraphic letters and the cardinality of $\setQ$ is $|\setQ|$. The operator $\mathbb{E}$ denotes expectation. 
\section{Prerequisites}

We now introduce the system model and MAP-JED optimization problem from which we derive a computationally efficient soft-output JED algorithm in \fref{sec:JED}.

\subsection{System Model}\label{sec:system model}
 
We focus on the uplink of a MU-MIMO communication system in which  $U$ single-antenna UEs transmit pilots and data to a BS equipped with $B$ antennas.
We assume a block-fading scenario with a coherence time of $K=T+D$ time slots; $T$ time slots are reserved for pilots and $D$ time slots are used for payload data.
The transmitted data matrix $\bS=[\bS_T,\bS_D]$ contains the pilots $\bS_T\in\complexset^{U\times T}$  and the transmit symbols $\bS_D\in\setQ^{U\times D}$ of all UEs, where $\setQ$ is the constellation set  (e.g., QPSK). 
In what follows, we consider a frequency-flat channel\footnote{Frequency selective channels can be transformed into parallel frequency-flat subcarriers by means of orthogonal frequency-division multiplexing.} with the following input-output relation~\cite{gesbert2003theory}:
\begin{align}\label{eq:mimo model}
\bY = \bH\bS + \bN.
\end{align} 
Here, $\bY \in \complexset^{B\times K}$ is the receive matrix containing all received symbols at the $B$ BS antennas over the $K$ time slots, $\bH\in\complexset^{B\times U}$ is the (unknown) channel matrix, and $\bN\in\complexset^{B\times K}$ models thermal noise with i.i.d.\ circularly-symmetric complex Gaussian entries and variance $\No$ per complex dimension.

\subsection{MAP-JED Optimization Problem}
 
We start by formulating the MAP-JED problem.
Our goal is to \emph{jointly} estimate the channel matrix and recover the most likely transmit symbols which requires us to assume priors for the channel matrix and the transmit symbols. For simplicity, we assume i.i.d.\ circularly-symmetric Gaussian entries in $\bH$ with entry-wise variance $E_h$ and equally likely transmit symbols.
These assumptions result in the MAP-JED problem
\begin{align}\label{eq:joint ML with H prior}
	\big\{\widehat \bH,\widehat\bS_D\big\} 
	& =  \argmin_{\substack{\bH\in\complexset^{B\times U}\\\bS_D\in\setQ^{U \times D}} } \normfro{\bY - \bH\bS}^2 + \lambda\normfro{\bH}^2
\end{align}
with $\lambda=\No/E_h$.\footnote{In \fref{sec:hypernetwork}, we let the hyper-network determine an appropriate choice of the parameter $\lambda^{(t)}$ for every iteration $t$, which enables us to apply our JED algorithm to channel matrices that are not necessarily i.i.d.\ Gaussian.}
For simplicity of exposition, we will directly work with the transmit symbol matrix $\bS$ instead of the pilot and data matrices $\bS_T$ and $\bS_D$, respectively. 

Since \fref{eq:joint ML with H prior} can be written as two nested optimization problems in the variables $\bS$ and $\bH$, we can first determine the optimal channel estimate $\widehat{\bH}$ given the transmit symbol matrix~$\bS$ and then find the optimal transmit symbol matrix $\widehat{\bS}$.
Since the optimization problem is quadratic in $\bH$, the optimal channel estimate $\widehat{\bH}$ has the following closed-form solution:
\begin{align} \label{eq:optchannelest}
    \widehat\bH =\bY\bS^H\bM^{-1},
\end{align}
where we use the auxiliary matrix $\bM = \bS\bS^H + \lambda\bI_{U}$.
We can now substitute $\widehat\bH$ into the objective of \fref{eq:joint ML with H prior} and perform algebraic simplifications, which leads to an equivalent MAP-JED problem that only depends on the transmit symbol matrix: 
\begin{align}\label{eq:opt1}
	\widehat{\bS} = \argmax\limits_{\bS\in\setQ^{U\times K}}\, \Tr\!\left[\bY^H\bY\bS^H\bM^{-1}\bS\right]\!.
\end{align}
After solving the MAP-JED problem in \fref{eq:opt1}, one can determine the optimal channel estimate by plugging $\widehat\bS$ into~\fref{eq:optchannelest}. 
Note that for $\lambda = 0$, the MAP-JED problem in  \fref{eq:opt1} reduces to the well-known maximum likelihood JED problem in \cite[Eq.~6]{xu2008exact}.
\section{S-JED: Soft-Output Joint Channel \\ Estimation and Data Detection}\label{sec:JED}

The combinatorial nature of solving \fref{eq:opt1} exactly would quickly result in prohibitive complexity  which necessitates approximate algorithms. 
Furthermore, even solving \fref{eq:opt1}  approximately would lead to a hard-output JED method that is unable to realize the full potential of coded data transmission. 
We now derive a computationally efficient algorithm to approximately solve \fref{eq:opt1} while being able to compute soft-output information. 
\subsection{Smoothening the MAP-JED Problem}\label{sec:convex relaxation}
The discrete nature of the constellation $\setQ$ is the culprit of preventing gradient-descent-like methods (and deep unfolding) to approximately solve \fref{eq:opt1}. We therefore propose to first relax the set  $\setQ$ to its convex hull, which is defined as \cite{castaneda2017vlsi}
\begin{align}
\setC = \textstyle \left\{\sum_{i=1}^{|\setQ|}\alpha_i s_i \mid (\alpha_i\in\reals_+,\forall i) \wedge \sum_{i=1}^{|\setQ|}\alpha_i =1 )\right\}\!.
\end{align}
Here,  $s_i$  is the $ i $th symbol in the constellation $ \setQ $. We can now replace the discrete constellation $\setQ$ in~\fref{eq:opt1} by the convex set $\setC$, which leads to the smoothened optimization problem 
\begin{align}\label{eq:opt2}
\widehat{\bS}^\text{sm} =  \argmax\limits_{\bS\in\setC^{U\times K}} \, \Tr\!\left[\bY^H\bY\bS^H\bM^{-1}\bS\right]\!.
\end{align}
The resulting smoothened optimization problem has a differentiable objective and a convex constraint, which permits the use of gradient-based methods and deep unfolding.

\subsection{Smoothened MAP-JED via Forward-Backward Splitting}
We now show how to use forward-backward splitting (FBS)~\cite{goldsteinField2016} to approximately solve \fref{eq:opt2}.  
FBS iteratively solves convex optimization problems of the form  
\begin{align} \label{eq:fbsproblem}
\hat\bms = \argmin_{\bms} f(\bms)+g(\bms),
\end{align}
where the function $f$ is differentiable and convex, and $ g $ is convex but not necessarily smooth or bounded.
By initializing FBS with $\bms^{(1)}$,  it solves the problem in~\fref{eq:fbsproblem} for the iterations $ t=1,2,\ldots $ until convergence by computing  
\begin{align}\label{eq:fbs alg}
\bms^{(t+1)} = \text{prox}_g(\bms^{ (t) }- \tau^{(t)} \nabla f( \bms^{ (t) } );\tau^{(t)} ),
\end{align}
where $ \nabla f(\bms) $ is the gradient of $ f(\bms) $ and $ \tau^{(t)}$ is a carefully-chosen step size at iteration $t$.
The proximal operator for~$ g(\bms) $ is defined as follows \cite{parikh2014proximal}:
\begin{align}\label{eq:proximal def}
\text{prox}_g(\bms;\tau) = \argmin_\bmz \, \tau g(\bmz) + \textstyle \frac{1}{2}\normtwo{\bmz-\bms}^2.
\end{align}
While FBS is able to exactly solve convex optimization problems, it can be used to approximately solve many non-convex problems~\cite{goldsteinField2016}, which will be described next.

For our MAP-JED problem, we define $f$ and $g$ in \fref{eq:fbsproblem} as 
\begin{align}
	f(\bS) = \Tr\!\left[\bY^H\bY\bS^H \bM^{-1}\bS\right] \quad \text{and} \quad 
	g(\bS)=  \chi_{\setC}(\bS),
\end{align}
where the indicator function $\chi_{\setC}(\bS)$
implements the convex constraint $\bS\in\setC^{U\times K}$ in \fref{eq:opt2}, 
and is zero if $\bS\in\setC$ and infinity otherwise. 
The gradient of the function $f$ in $\bS$ is given by  
\begin{align}\label{eq:nabla f}
	\nabla f(\bS) &= \bM^{-1}\bS\bY^H\bY (\bI_{U}-\bS^H\bM^{-1}\bS).
\end{align}
Due to space constraints, the derivation of the gradient will be shown in the future journal version of the paper \cite{songSJEDfuture}.
The proximal operator in \fref{eq:proximal def} for $\bS$ is given by 
\begin{align}
&\text{prox}_{g}(\bS;\tau^{(t)})  
= \argmin_{\bX\in\setC^{U\times K}} \textstyle \frac{1}{2}\normfro{\bX-\bS}^2, \label{eq: NC proximal S}
\end{align}
where we move the indicator function in $g(\bS)$ back to the constraint. 
This problem has closed-form expressions for most constellation sets, e.g., for QPSK the projection operator is 
\begin{align}\label{eq:inf proximal}
\text{proj} _\setC(\Re\{S_{i,j}\}) &= \min\{\max\{\abs{\Re\{S_{i,j}\}},-\alpha\},\alpha\}\\
\text{proj} _\setC(\Im\{S_{i,j}\}) &= \min\{\max\{\abs{\Im\{S_{i,j}\}},-\alpha\},\alpha\},
\end{align}
where $ \alpha =\frac{1}{\sqrt{2}}$ for the set $\setQ=\big\{\pm\frac{1}{\sqrt{2}}\pm \frac{1}{\sqrt{2}}j\big\}$.

This FBS algorithm efficiently (and approximately) solves the smoothened MAP-JED problem in~\fref{eq:opt2}, but 
ignores the discrete nature of the constellation set and is unable to compute soft-outputs. Sections~\ref{sec:posterior mean} and \ref{sec:approxsoftoutputtrick} address these issues. 

\subsection{Exploiting the Constellation with the PME}\label{sec:posterior mean}
In order to exploit the constellation set $\setQ$, we model the iterations of FBS after evaluating the gradient step  
\begin{align} \label{eq:gradientstep}
\bX^{(t)} = \bS^{ (t) }- \tau^{(t)} \nabla f( \bS^{ (t) })
\end{align}
as follows: 
\begin{align}\label{eq:post equalized additive model}
\bX^{(t)}= \bS + \bE^{(t)}.
\end{align}
Here, $\bS$ is the (unknown) true transmitted data matrix and~$\bE^{(t)}$ models estimation errors on these per-iteration estimates. By assuming that the distribution of~$\bE^{(t)}$ is known, one can replace the projection onto the convex hull $\setC$ in \fref{eq: NC proximal S} by the entry-wise posterior mean estimate (PME)
\begin{align} \label{eq:PME}
S_{u,k}^{(t+1)} = \Exop[ S_{u,k}| X_{u,k}^{(t)}]
\end{align}
with $u=1,\ldots,U$ and $k=1,\ldots,K$. We emphasize that the PME depends on the prior distribution (which is given by the constellation set $\setQ$) and the statistics of $\bE^{(t)}$. 
We have the following prior distribution on the transmitted data:
\begin{align}
p(S_{u,k}) = \frac{1}{|\setQ|}\sum_{i=1}^{|\setQ|} \delta(S_{u,k}-s_i).
\end{align}
Here, $s_i$ is the $i$th symbol in the constellation $\setQ$ and $\delta$ is the Dirac delta function. 
By assuming that the estimation errors~$E^{(t)}_{u,k}$ are circularly-symmetric complex Gaussian with variance $\nu_{u,k}^{(t)}$, the PME in \fref{eq:PME} has a closed form (see, e.g.,~\cite{jeonOptimality2015}) and one can replace the proximal operator in \fref{eq: NC proximal S} with the PME in \fref{eq:PME}.
This step requires knowledge of the per-iteration estimation error variances $\nu_{u,k}^{(t)}$, $\forall u,k$, which are difficult to obtain in practice. However, as shown in \fref{sec:hypernetwork}, we can use a hyper-network to determine these variances.

\subsection{Approximate PME and Soft-Output Generation}
\label{sec:approxsoftoutputtrick}
In order to extract soft-outputs, we build upon the approximate PME put forward in \cite{jeonMismatched2020,jeon3542019}. 
The key idea is to replace~\fref{eq:PME} by an approximate three-step procedure that (i) converts the per-iteration estimates in \fref{eq:gradientstep} into soft-outputs in the form of log-likelihood ratios (LLRs) for every transmitted bit, (ii) transforms these LLR values into probabilities, and (iii) converts these probabilities back into the soft-symbol estimates. 
We now summarize this approach for QPSK modulation. 

\subsubsection*{Step (i)}
Assume that the per-iteration estimation errors are circularly-symmetric complex Gaussian with variances $\nu_{u,k}^{(t)}$, $\forall u,k$. 
Then, the LLRs for the two bits that map to QPSK symbols are given by \cite[Tbl. 4.3]{fateh2009vlsi} 
\begin{align}\label{eq:LLR1}
	L_{1,u,k}^{(t)} = \frac{4\Re \{X_{u,k}^{(t)}\}}{\nu_{u,k}^{(t)}} \quad \text{and} \quad L_{2,u,k}^{(t)} = \frac{4\Im \{X_{u,k}^{(t)}\}}{\nu_{u,k}^{(t)}}.  
\end{align}

\subsubsection*{Step (ii)}
We convert the LLRs in~\fref{eq:LLR1} into probabilities as follows \cite[Eq.~3.6]{fateh2009vlsi}:
\begin{align} \label{eq:probabilities}
P_{b,u,k}^{(t)} & = \frac{1}{2}\!\left(1+\tanh\!\left(\frac{L_{b,u,k}^{(t)}}{2}\right)\!\right)\!, \quad b\in\{1,2\},
\end{align}
which express the probabilities of the $b$th bit that map to the symbol~$S_{u,k}$ (e.g., using Gray mapping) being $1$.

\subsubsection*{Step (iii)}
We use the probabilities in \fref{eq:probabilities} to compute symbol estimates as follows \cite[App.~A.4]{fateh2009vlsi}:
\begin{align}
\Re\{S_{u,k}^{(t+1)}\} &= {\frac{1}{\sqrt{2}}}(2P_{1,u,k}^{(t)}-1) \label{eq:approximate PME 1} \\
\Im\{S_{u,k}^{(t+1)}\} &= {\frac{1}{\sqrt{2}}}(2P_{2,u,k}^{(t)}-1)\label{eq:approximate PME 2},
\end{align}
where $S_{u,k}^{(t+1)}$ approximates the PME output in \fref{eq:PME}.

We emphasize that we are using this three-step procedure instead of the PME in \fref{eq:PME} for three reasons: 
First, in Step~(i) we calculate LLR values for every transmitted bit in each iteration, which provides our algorithm with soft-output data detection capabilities. 
Second, in Step~(ii) we obtain probabilities for every transmitted bit in each iteration, which will be key in learning the parameters of our soft-output JED algorithm (see \fref{sec:hypernetwork training}). 
Third, this procedure was shown in~\cite{jeon3542019,jeonMismatched2020} to be less complex and numerically more stable than evaluating the exact PME in \fref{eq:PME}.
\begin{figure*}[tp]
\centering
\includegraphics[width=0.99\textwidth]{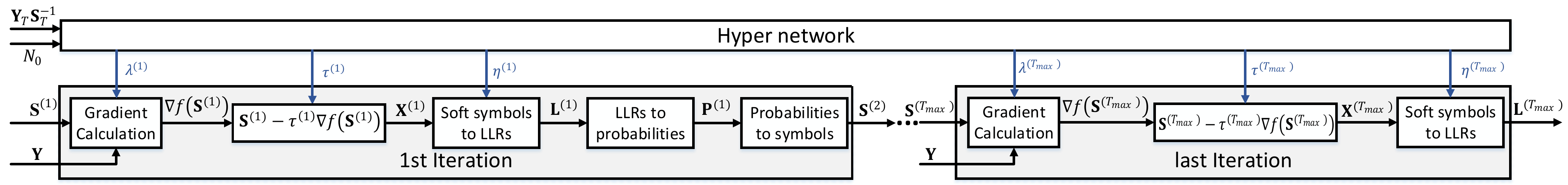}
\caption{ 
Block diagram of the deep unfolded soft-output JED and the hyper-network.
The unfolded algorithm consists of $T_\text{max}$ layers. 
Each layer takes in soft-symbols from the preceding layer and outputs new soft-symbols; the last layer outputs only LLR values.
The hyper-network takes in the LS channel estimate $\widehat\bH^{\text{LS}}$ and noise variance $\No$ in order to produce the parameters  step sizes $\tau^{(t)}$, regularization parameters $\lambda^{(t)}$, and normalized error variances $\eta^{(t)}_{u}$. }
\label{fig:overall}
\vspace{-0.2cm}
\end{figure*}
\section{Deep Unfolding with a Hyper-Network}\label{sec:hypernetwork}
We now explain our deep unfolding strategy for the soft-output JED algorithm and how to train the algorithm parameters. 
Due to space constraints, we focus on QPSK only---the general case will be presented in~\cite{songSJEDfuture}.

\subsection{Deep-Unfolding Architecture}\label{sec:unfolding algorithm}
In order to determine the algorithm parameters, we use an emerging paradigm known as deep unfolding \cite{hersheyDeep2014,balatsoukas-stimming19a,mongaAlgorithm2021} which we combine with a hyper-network that provides these parameters based on estimated CSI \cite{goutayDeep2020}.
The idea of deep unfolding is to unfold an iterative algorithm into $T_\text{max}$ layers (one for every iteration) and use tools of deep learning to determine an optimal set of the algorithm's parameters in every iteration (layer) $t=1,\ldots,T_\text{max}$. Instead of hard-coding these parameters after training, we train a hyper-network that generates these algorithm parameters dependent on CSI. 
 
The hyper-network, the unfolded algorithm (which consists of $T_\text{max}$~layers, each representing a JED iteration), and the parameters are shown in \fref{fig:overall}.
At layer~$t$, the input $\bS^{(t)}$ is first updated by a gradient descent step in~\fref{eq:gradientstep} to obtain the symbol estimates~$\bX^{(t)}$.
Then, the three-step procedure to approximate the PME as described in \fref{sec:approxsoftoutputtrick} is performed to obtain the next iterate $\bS^{(t+1)}$ as in~\fref{eq:approximate PME 1} and \fref{eq:approximate PME 2}.
We note that the last layer $t=T_\text{max}$ only requires the LLR outputs in \fref{eq:LLR1}.

Our unfolded architecture requires several algorithm parameters, which are generated by a hyper-network. 
Specifically, for each iteration (layer) $t=1,\ldots,T_\text{max}$, we require the per-iteration step size~$\tau^{(t)}$ and the estimation error variances $\nu_{u,k}^{(t)}$ for $u=1,\ldots,U$ and $k=1,\ldots,K$.
To reduce the amount of parameters per iteration, we assume that the estimation error variances are fixed with respect to the time slot $k$, i.e., we only require $\nu_{u}^{(t)}$ and use the same variance for all time slots. 
We note that the hyper-network does not generate~$\nu_{u}^{(t)}$, but rather a normalized version $\eta_{u}^{(t)} = {\No}/{\nu_{u}^{(t)}}$ to account for large variations in $\No$. 
We also require the parameter~$\lambda$ in~\fref{eq:joint ML with H prior}; instead of using the same parameter $\lambda$ for all iterations $t=1,\ldots,T_\text{max}$, each layer uses a different parameter $\lambda^{(t)}$. 

The inputs to the hyper-network are the vectorized least-squared channel estimate $\widehat\bH^{\text{LS}} = \bY_T\bS_T^{-1}$ of the pilot phase ($\bY_T$ contains the first $T$ columns of the matrix $\bY$) and the  noise variance~$\No$. 
The hyper-network itself consists of five dense layers with rectified linear unit (ReLU) activations in each layer except for the last one, which uses an absolute value activation to generate non-negative parameters.

\subsection{Hyper-Network Training}\label{sec:hypernetwork training}
In order to train the hyper-network, we leverage the soft-output capabilities of our algorithm.
Specifically, since our JED algorithm computes probabilities for the transmitted bits~\fref{eq:probabilities}, we can train the hyper-network using the outputs in the last iteration $t=T_\text{max}$ using the widely-used binary cross entropy (BCE) loss, which is defined as follows: 
\begin{align}\label{eq:bce canonical}
 H(b_i,p(b_i)) =  b_i \log(p(b_i)) + (1-b_i)\log(1-p(b_i)).
\end{align}
Here, $b_i\in\{0,1\}$ is the label of the $i$th bit and $p(b_i)$ is the predicted probability of this bit being $1$.
In our case, we utilize the probabilities $P_{b',u,k}^{(T_\text{max})}$ in \fref{eq:probabilities} for every transmitted bit $b_{b',u,k}$, where $b'=1,2$ is the bit index, $u=1,\ldots,U$ the UE index, and  $k=1,\ldots,K$ the time slot index, calculated in the last iteration $t=T_\text{max}$. 
Hence, we define the following average BCE loss over all of these probabilities
\begin{align}\label{eq:loss function}
L = \frac{1}{2UK}\sum_{b'=1}^{2}\sum_{u=1}^{U}\sum_{k=1}^{K} H\!\left(b_{b',u,k},P_{b',u,k}^{(T_\text{max})}\right)\!,
\end{align}
which we use to train the hyper-network parameters. We learn only a single hyper-network for all signal-to-noise-ratio (SNR) values, which is in stark contrast to the common approach of using a different hyper-network for every SNR. 
 
\section{Simulation Results}

We now demonstrate the efficacy our soft-output JED algorithm and compare it to baseline algorithms. We first detail the system setup and then show simulation results. 

\subsection{System Setup}\label{sec:simulation setup}

We simulate a  MU-MIMO system as described in \fref{sec:system model} with $B=8$ BS antennas and $U=4$ single-antenna UEs transmitting QPSK symbols for $K=244$ time slots. The UEs transmit orthogonal pilots in  $\bS_T$ from a $4\times4$ Hadamard matrix.
The channel matrices are modelled as Rayleigh fading with i.i.d. complex standard Gaussian entries.
We consider per-UE coding with a  rate-1/2 low-density parity-check (LDPC) code as in IEEE 802.11n~\cite{wlan80211nPHY} with a block-length of $480$ bits; for LDPC decoding, we use a sum-product and layered decoding algorithm with $10$ iterations.
The hyper-network is trained using an NVIDIA GTX1080 with $1\,$M transmissions and batch size of $1\,$k.
We use Monte-Carlo simulations to extract the coded packet error rate (PER), uncoded bit error rate (BER), and BCE as in \fref{eq:loss function}.
We run $T_\text{max}=10$ iterations of our soft-output JED algorithm (called ``S-JED'').

\subsection{Baseline Algorithms}
In order to evaluate the effectiveness of our S-JED algorithm, we simulate the SIMO lower bound, which cancels  MU interference with perfect CSI in a genie-aided fashion~\cite{zhang2006non}.
We also compare our algorithm to conventional methods that separate channel estimation from soft-output data detection. 
For such methods, we simulate a SIMO lower bound with estimated CSI (called ``SIMO (est. CSI)''), where we use a least-squares channel estimator to compute~$\widehat{\bH}^\text{LS}$. 
We also compare S-JED to the widely used soft-output linear minimum mean-square error (L-MMSE) equalizer \cite{seethalerEfficient2004,fateh2009vlsi} and the max-log optimal single-tree-search sphere decoder (STS-SD)~\cite{studer08a}.
\subsection{Simulation Results}
\begin{figure*}[tp]
\centering
\subfigure[]{\includegraphics[width=0.31\textwidth]{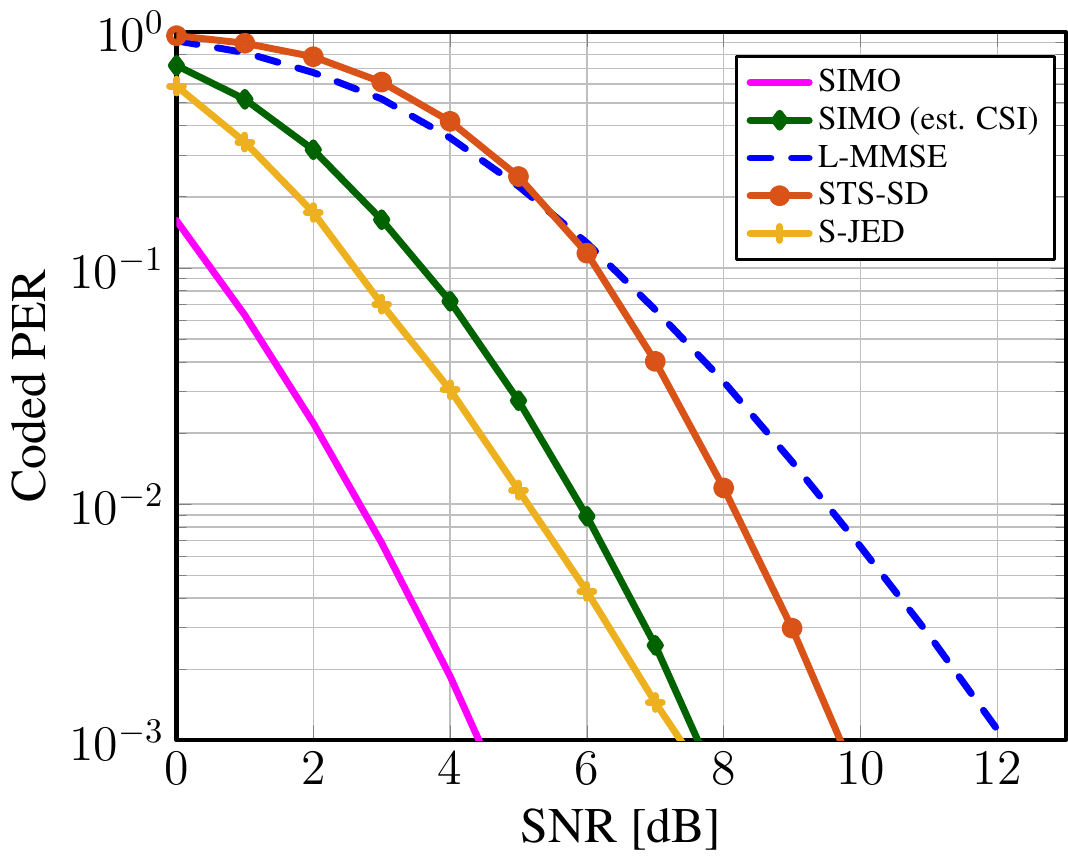}}
\hspace{0.2cm}
\subfigure[]{\includegraphics[width=0.31\textwidth]{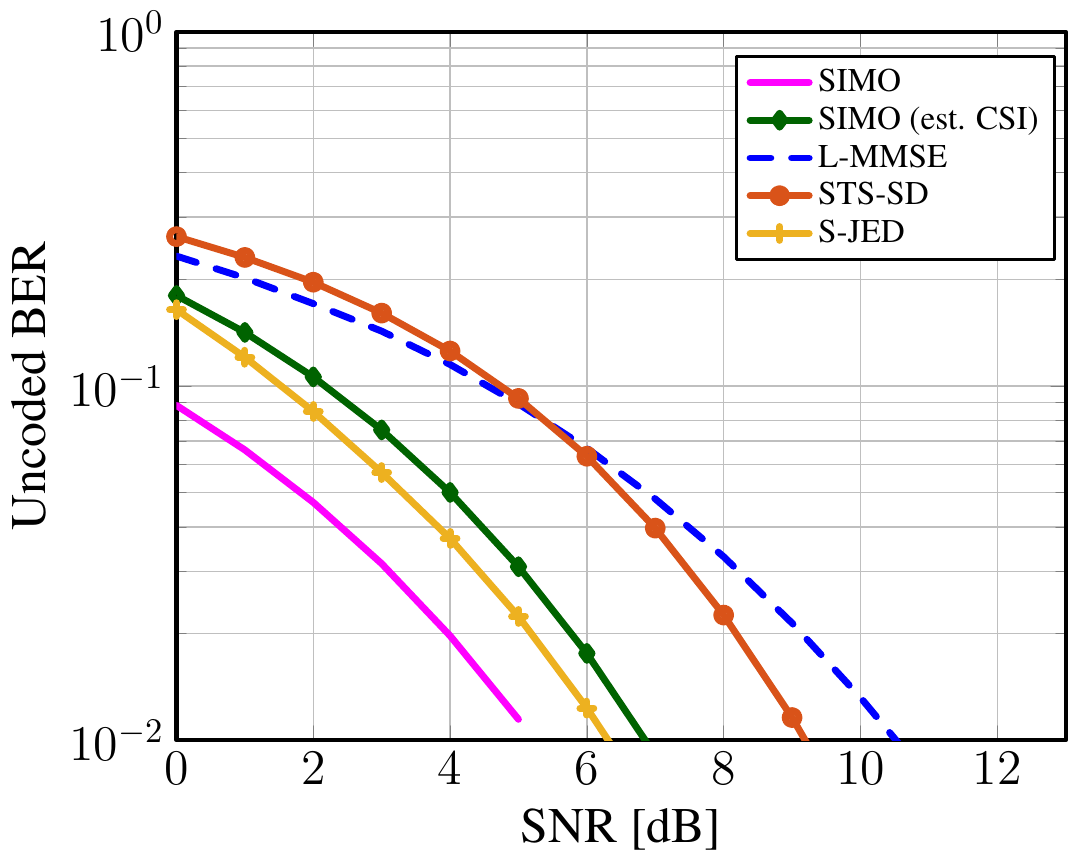}}
\hspace{0.2cm}
\subfigure[]{\includegraphics[width=0.31\textwidth]{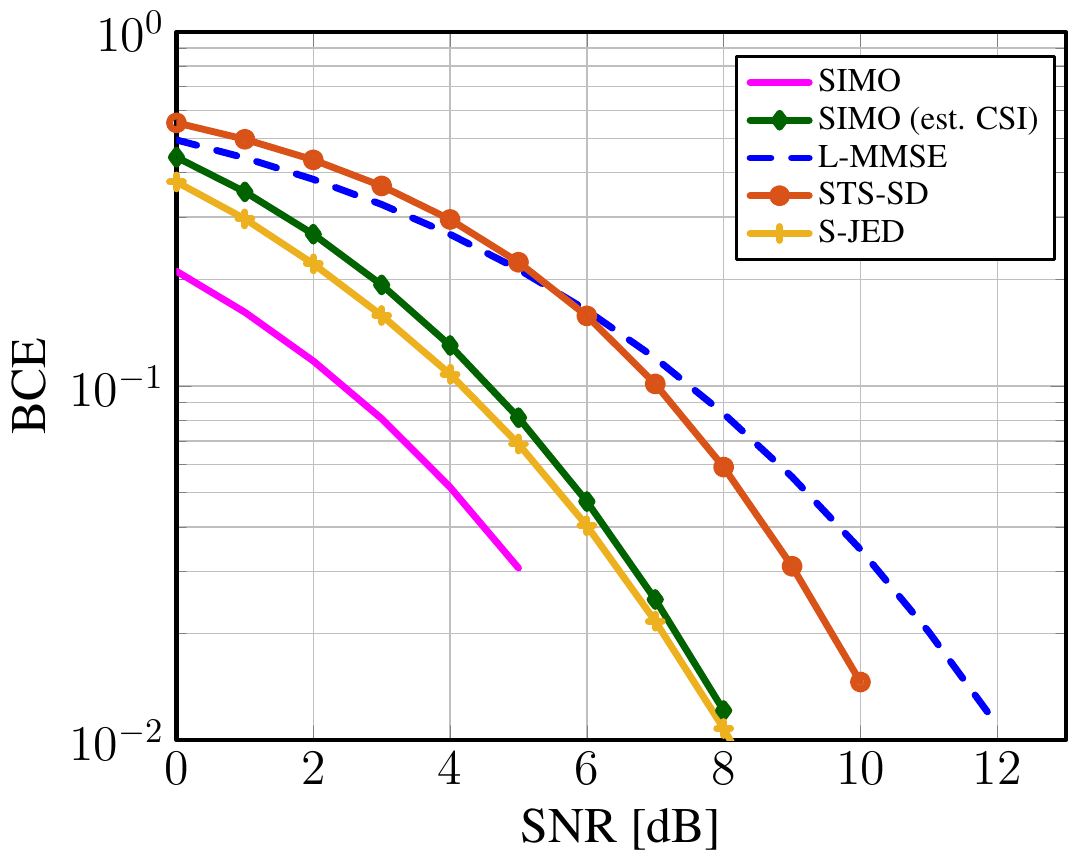}}\\
\caption{Coded PER (a), uncoded BER (b), and BCE (c) performance for a $B=8$ BS antenna, $U=4$ UE MU-MIMO system with  transmitting QPSK for $K = 240$ time slots.
The proposed soft-output JED (S-JED) algorithm approaches the SIMO lower bound and outperforms the SIMO bound with estimated CSI as well as the max-log optimal soft-output STS-SD and the widely used L-MMSE equalizer which separate channel estimation from  data detection.}
\label{fig:8x4 numerical results}
\vspace{-0.2cm}
\end{figure*}

Figure~\ref{fig:8x4 numerical results} shows our simulation results. 
In \fref{fig:8x4 numerical results}(a), we see that S-JED approaches the SIMO lower bound by less than $3\,$dB at a coded PER of $0.1$\% and outperforms the SIMO lower bound that uses estimated CSI. 
S-JED significantly outperforms the max-log optimal soft-output STS-SD algorithm and the widely used L-MMSE equalizer, which both separate channel estimation from detection, by $2$\,dB and $4$\,dB, respectively.
In \fref{fig:8x4 numerical results}(b), we see that the uncoded BER results behave similarly. 
The results in \fref{fig:8x4 numerical results}(c) demonstrate that the BCE accurately characterizes the performance of all methods, as the order between algorithms is preserved with respect to coded PER and uncoded BER---this implies that the BCE loss in \fref{eq:loss function} is well suited to train soft-output data detectors.

\section{Conclusions}
  
We have proposed a novel soft-output joint channel estimation and data detection (S-JED) algorithm for MU-MIMO systems.
Our method formulates a maximum a-posteriori (MAP) optimization problem and computes approximate LLR values in every iteration. 
The algorithm parameters are generated by a hyper-network, which is trained using deep unfolding and a BCE cost function. 
Simulation results have shown that the proposed S-JED algorithm with only 10 iterations significantly outperforms the max-log optimal STS-SD and L-MMSE equalizer, which both separate channel estimation from soft-output data detection. 

There are many avenues for future work. Our future journal paper in~\cite{songSJEDfuture} will include missing derivations, a complexity comparison, and apply S-JED to higher-order modulation schemes as well as to other channel models.

\balance
\bibliographystyle{IEEEtran}
\bibliography{bibs/VIPabbrv,bibs/publishers,bibs/confs-jrnls,bibs/vipbib_hc}
\balance
\end{document}